\documentclass[10pt,letterpaper,twocolumn]{article} 

\usepackage{ol2}
\usepackage{amsmath}

\begin{document}

\twocolumn[ 

\title{Moment analysis of focus-diverse point spread functions 
for modal wavefront sensing of uniformly illuminated circular-pupil systems} 

\author{Hanshin Lee$^*$} 
\address{
McDonald Observatory, University of Texas at Austin, 1 University
Station C1402, Austin, TX, 78712, USA\\ 
$^*$Corresponding author: lee@astro.as.utexas.edu
}

\begin{abstract}
A new concept of using focus-diverse point spread functions (PSFs) 
for modal wavefront sensing (WFS) is explored. This is based on 
relatively straightforward image moment analysis of measured PSFs, 
which differentiates it from other focal-plane wavefront sensing techniques (FPWFS). 
The presented geometric analysis shows that the image moments are 
non-linear functions of wave aberration coefficients, but notes that 
focus-diversity (FD) essentially decouples the coefficients of 
interest from others, resulting in a set of linear equations whose
solution corresponds to modal coefficient estimates.
The presented proof-of-concept simulations suggest the potential
of the concept in WFS with strongly aberrated high SNR objects 
in particular.\\ 
\end{abstract}


] 
\noindent 
Optical wavefront is aberrated during its passage through
physical systems due to perturbations like turbulence in 
Earth atmosphere or misalignment of telescopes. 
Useful understanding of the observed behaviours of the 
systems can be obtained by sensing the aberration 
coefficients of the wavefront. Typically, a wavefront is 
measured at the pupil of an imaging system, as in 
Shack-Hartmann sensors (SHS) where wavefront slope is 
measured and then wave aberration is determined from the 
measurement\cite{Hardy}. It can be efficient and accurate, 
but requires a separete relay for pupil re-imaging with 
additional optical components.

In some cases, one wishes to sense wavefront across the 
field of view of an instrument by using the built-in 
components without separate relay components. 
Such desires can be satisfied by applying FPWFS techniques to 
focus-diverse images. Curvature sensing (CS) is a FPWFS method 
where the intensity difference between two extra-focal images 
are used to determine wavefront\cite{RoddierCS}. The CS signal 
is directly related to the shape of a membrane deformable mirror, 
enabling efficient control of the wavefront compensator\cite{Hardy}.
In Phase retrieval, another FPWFS method, wavefront is optimized
until its synthetic focus-diverse images closely match the 
measured ones\cite{Fineup}. It can be accurate and flexible, 
but may be less efficient than others due to a potentially 
large number-crunching during optimization. However, progress 
has been made to address this issue\cite{Dolne,Gonsalves,LIFT}.

In this Letter, the idea of a new type of FPWFS technique is explored. 
Like others, it utilizes focus-diverse PSFs, but differs, in 
that aberrations are sensed via image moment
analysis of the PSFs. The presented geometric analysis shows 
that the moments are non-linear functions of wave aberration 
coefficients, but notes that FD essentially decouples the 
coefficients of interest, leading to linear equations whose 
solution corresponds to modal coefficient estimates. It is 
argued that this approach has potential in WFS as 
shown in the presented simulations. 
Noll's notations are adopted throughout\cite{Noll}. 

\textit{Geometrically}, aberrated systems spread rays around 
and blur their PSFs, but the blur shape uniquely differs 
depending on wave aberrations, e.g. comet-like PSF blur by coma 
(see pp. 90 in\cite{Mahajan} for more examples). 
Such geometric relations can be explained by the connection 
between ray and wave aberrations.
Suppose that a wavefront $\Phi$ on the unit-disk pupil $\Omega$ 
is given as, 
\begin{align}
	\Phi = a_1\,Z_1 + a_2\,Z_2 + a_3\,Z_3 + \cdots  + a_M\,Z_M,
	\label{ZinW}
\end{align}
\noindent where $a_i$ are the wave aberration coefficients 
and $Z_i$ are the Zernike polynomials in the pupil coordinates ($x,y$). 
The ray coordinates ($X,Y$) at the focus are given by differentiating 
$\Phi$ by $x$ and $y$, respectively.
\begin{align}
	X=X_0+2F\,\frac{\partial \Phi}{\partial x},\;\;
	Y=Y_0+2F\,\frac{\partial \Phi}{\partial y},
        \label{eqn2}
\end{align}
\noindent where ($X_0,Y_0$) are the ideal image coordinates and $F$ 
is the focal ratio. The derivative terms are given as,
\begin{align}
	\begin{array}{c}
	2F\;\partial\Phi/\partial x = \vec{a}^{T}\,\mathbf{\Gamma}^x\,\vec{Z} 
	= {(\vec{a}{'}^{x})}^T \,\vec{Z}\\
	2F\;\partial\Phi/\partial y = \vec{a}^{T}\,\mathbf{\Gamma}^y\,\vec{Z} 
	= {(\vec{a}{'}^{y})}^T \,\vec{Z}
	\end{array},
	\label{waveslope}
\end{align}
\noindent where $\vec{a}{'}^{x}$ and $\vec{a}{'}^{y}$ are slope coefficient 
vectors and $\vec{Z}$ is the vector of $Z_i$. 
Noll noted the derivatives of $Z_i$ can be given in terms of 
$Z_i$ through the conversion matrices $\mathbf{\Gamma}^x$ and 
$\mathbf{\Gamma}^y$\cite{Noll}.
$a{'}_i^{x}$ and $a{'}_i^{y}$ are related to $a_i$, e.g. for $M$=10
\begin{align}
	a{'}_1^x=&2a_2+2\sqrt{2}a_8,\;\;\;\;\;
	a{'}_1^y=2a_3+2\sqrt{2}a_7, \nonumber \\
	a{'}_2^x=&2\sqrt{3}a_4+\sqrt{6}a_6,\;\;
	a{'}_2^y=\sqrt{5}a_5+\sqrt{10}a_{13}, \nonumber \\
	a{'}_3^x=&\sqrt{5}a_5+\sqrt{10}a_{13},\;
	a{'}_3^y=2\sqrt{3}a_4-\sqrt{6}a_6\nonumber \\
	a{'}_4^x=&2\sqrt{6} a_8,\;\;\;\;\;\;\;\;\;\;\;\;\;\;\;\;\;
	a{'}_4^y=2\sqrt{6} a_7, \nonumber \\
	a{'}_5^x=&2\sqrt{3}a_7+2\sqrt{3}a_9,\;
	a{'}_5^y=2\sqrt{3}a_8-2\sqrt{3}a_{10},\nonumber \\
	a{'}_6^x=&2\sqrt{3}a_8+2\sqrt{3}a_{10},
	a{'}_6^y=-2\sqrt{3}a_7+2\sqrt{3}a_{9}.
	\label{slope}
\end{align}

One way to quantify the PSF shape is computing
its $k$th moment of order $k=n+m \geq 1$, that is given as,
\begin{align}
	\mu_{nm}=S^{-1}\int_{\Omega} I\,(X-X_0)^n\,(Y-Y_0)^m d\Omega,
	\label{moments}
\end{align}
\noindent where $I$ is the pupil illumination and $S$ is its 
integral over $\Omega$. Here, the moment is computed using the pupil plane 
quantities (i.e. $I$, $X$, and $Y$), whereas this is \textit{usually} done using
focal plane quantities. But, note that the focal plane PSF is 
essentially determined by the pupil plane quantities. 
Thus, Eq.~\ref{moments} should lead to the moment as computed in the usual way. 

With $I=1$ over $\Omega$, $S=\pi$ and the centroids become $\mu_{10}=a{'}_1^x$ and 
$\mu_{01}=a{'}_1^y$. Apparently, $\mu_{nm}$ with $k\geq2$ is non-linear in $a{'}_i^x$ 
or $a{'}_i^y$, but only some slope coefficients (say $b_i$) are coupled to 
$(a{'}_2^x)^\alpha (a{'}_3^y)^\beta$ 
where $\alpha+\beta+1=k$. Note that $a{'}_2^x$ and $a{'}_3^y$ are the only 
coefficients affected by defocus ($a_4$), thus differentiating these with respect 
to $a_4$ by $k$-1 times decouples $b_i$ from $(a{'}_2^x)^\alpha (a{'}_3^y)^\beta$.
For instance, the derivatives of $\mu_{nm}$ at $k$=2 (Eq.~\ref{moment2}),
\begin{align}
\mu_{20}= \sum_{i=2}^{M} (a{'}_i^x)^2,{\mu}_{11}=\sum_{i=2}^{M}\,a{'}_i^x a{'}_i^y,
	\mu_{02}= \sum_{i=2}^{M}\,(a{'}_i^y)^2, 
	\label{moment2}
\end{align}
\noindent 
are the followings.
\begin{align}
	\frac{\partial{\mu}_{20}}{\partial a_4} = 4\sqrt{3} a{'}_2^x,
	\frac{\partial{\mu}_{11}}{\partial a_4} = 4\sqrt{3} a{'}_3^x,
	\frac{\partial{\mu}_{02}}{\partial a_4} = 4\sqrt{3} a{'}_3^y.
	\label{dervmoment2}
\end{align} 
\noindent Note $a{'}_3^x = a{'}_2^y$ from Eq.~\ref{slope}. Another example is
for $k=3$, where the 2nd partial derivatives of ${\mu}_{nm}$ are
given as,
\begin{flalign}
	\frac{\partial^2{\mu}_{30}}{\partial{a^2_4}}&=c(a{'}_4^x+\sqrt{2}a{'}_6^x),
	\frac{\partial^2{\mu}_{12}}{\partial{a^2_4}}=d(5a{'}_4^x-3\sqrt{2}a{'}_6^x), \nonumber \\
	\frac{\partial^2{\mu}_{03}}{\partial{a^2_4}}&=c(a{'}_4^y-\sqrt{2}a{'}_6^y),
	\frac{\partial^2{\mu}_{21}}{\partial{a^2_4}}=d(5a{'}_4^y+3\sqrt{2}a{'}_6^y), \nonumber \\
	\mathrm{with}&\;a{'}_5^x=\sqrt{2}a{'}_4^y+a{'}_6^y,\;a{'}_5^y=\sqrt{2}a{'}_4^x+a{'}_6^x,
\end{flalign}
\noindent where $c=12\sqrt{3}$ and $d=4\sqrt{3}$. Solving the above 
equations produces the slope coefficients. 
In a similar way, the $k$-1th partial derivatives of the $k$th moments
can be used to estimate the slope coefficients at the $k$-1th order. The
slope estimates can then be used in Eq.~\ref{waveslope} to determine 
$a_i$ of radial order up to $k$. 

What does it mean to have the derivative of moment in reality then? It
is simply \textit{FD}. 
Through-focusing is one way to achieve it, where the PSF is recorded at $N$ 
different intra- or extra-focal planes and the image moments of the PSFs are
measured. This obviously leads to $\mu_{nm}$ that is a function of $\Delta a_4$. 
Note $N$=$k$ and $\Delta a_4$=$\Delta L/(16\sqrt{3}F^2)$ 
in the unit of focus perturbation $\Delta L$. 
Finally, fitting a polynomial of degree $k$-1 to the measured $\mu_{nm}$
results in the $k$-1th coefficient that corresponds to the $k$-1th
derivative of $\mu_{nm}$ with respect to $a_4$. In summary, one can obtain the
dependence of $\mu_{nm}$ on $a{'}_i^x$, $a{'}_i^y$ from
Eq.\ref{eqn2},~\ref{waveslope},~\ref{slope} and the $k$-1th derivative
of $\mu_{nm}$ with respect to $a_4$ leads to simple expressions
allowing to estimate $a_i$. To experimentally measure the derivative,
the PSF should be recorded at $k$ different image planes or FDs.

As a proof-of-concept, I demonstrate a simulation where non-zero  
$a_i$ with $i=2,3,4,\cdots,10$ are estimated by using three through-focus 
images sampled at -1mm, 0mm, and 1mm ($\Delta a_4$$\sim$$\pm0.7\lambda$) from 
the ideal focus of a f/10 system with 2.7m aperture (plate scale $\sim 7.6''/mm$). 
A standard FFT method was used for synthesising polychromatic PSFs\cite{Goodman}.
\begin{figure}[htpb]
	\centering 
	\includegraphics[width=8.0cm]{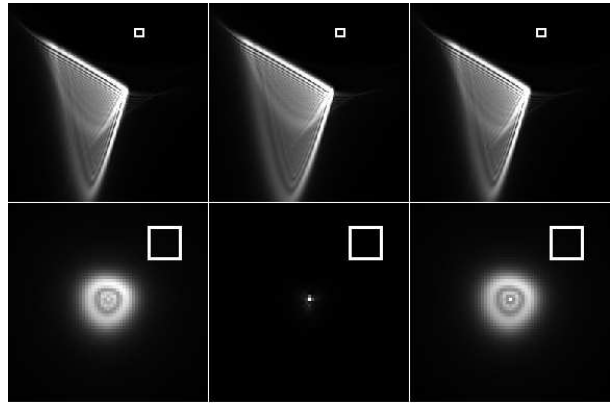}
	\caption{\footnotesize{Polychromatic through-focus PSFs on 512$^2$ grid
	 with uniform spectral weight at 11 wavelengths between 514nm and 614nm, 
	 Top: Initial, Bottom: After correction, Left: -1mm before focus, 
	 Middle: at focus, Right: +1mm after focus. 50$\mu m$ squares overlaid 
	 for size comparison. Images in square-root scale.}}
	\label{Figure1}
\end{figure}

The initial wavefront error is quite substantial amounting to 5.7$\lambda$ 
in rms at $\lambda$=514nm, making the through-focusing effect in the PSF 
shape barely noticeable. The pixel size is 5$\mu m$.
The estimated coefficients are compared to the input values in
Table~\ref{Table1}.
The estimated slope coefficients ($\hat{a}{'}_i^x$, $\hat{a}{'}_i^y$) 
match the true values (${a}{'}_i^x$, ${a}{'}_i^y$) with sub-pixel 
accuracy. The maximum difference between the estimated ($\hat{a}_i$) 
and the true ($a_i$) wave coefficients is 0.032$\lambda$ in $a_{9}$.
The PSFs after applying $\hat{a}_i$ as corrections are in the bottom row of Figure~\ref{Figure1}.
\begin{table}
  \centering
  \footnotesize
  \caption{\footnotesize{Slope (in pixel) and wave (in $\lambda$) coefficients in Figure~\ref{Figure1}.}}
  \begin{tabular}{c|c|c|c|c|c|c} \hline\hline
    $i^k$ & $\hat{a}{'}_i^x$ &  ${a}{'}_i^x$ & $\hat{a}{'}_i^y$ & $a{'}_i^y$ 
    & $\hat{a}_i$ & $a_i$\\ \hline
    1$^0$ &   -11.926 &   -11.959 &   -1.078  &  -1.144   &  --    &  --    \\ \hline
    2$^1$ &   -27.885 &   -27.986 &    7.956  &   7.968   & -1.198 & -1.185\\
    3$^1$ &     7.956 &     7.968 &  -30.874  &  -31.015   &  1.612  &  1.636\\ \hline
    4$^2$ &   -12.351 &   -12.505 &  -13.033  &  -13.316  & -4.241  & -4.241\\
    5$^2$ &     1.823 &     1.766 &  -18.399  &  -18.601  &  1.624 &  1.626\\
    6$^2$ &     0.932 &     0.917 &   20.255  &   20.598  &  0.305  &  0.309\\ \hline
    7$^3$ &     -- &     -- &    --  &   --  & -1.330  & -1.359\\
    8$^3$ &     -- &     -- &    --  &   --   & -1.261 & -1.276\\
    9$^3$ &     -- &     -- &    --  &   --   &  1.593  &  1.614\\
    10$^3$ &     -- &     -- &    --  &   --   &  1.392 &  1.408\\
\hline\hline
  \end{tabular}
  \label{Table1}
\end{table}

\begin{figure}[htpb]
	\centering 
	\includegraphics[width=8.0cm]{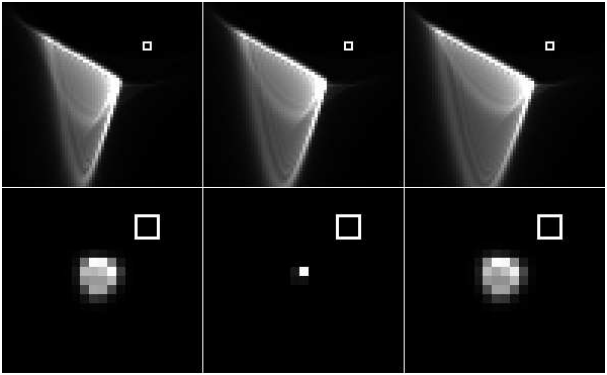}
	\caption{\footnotesize{Polychromatic through-focus PSFs with 4x4 binning, Top: Initial, 
	Bottom: After correction, Left: -1mm before focus, Middle: at focus, 
	Right: +1mm after focus. 
	50$\mu m$ squares overlaid for comparison. 
	Images in square-root scale.}}
	\label{Figure2}
\end{figure}

In the next demonstration, the same images were binned by a factor 
of 4 (Figure~\ref{Figure2}). The same moment analysis was applied to 
the binned images and the resultant wave and slope aberration 
coefficients are compared to the true values in Table~\ref{Table2}. 
The binning effectively reduces the spatial resolution of the images.
This leads to a fewer pixels used in the moment analysis and thus reduces
the moment measurement accuracy. To some extent, binning is equivalent to
low-pass filtering, washing out small-scale diffraction structures that
may not be accounted for by the geometric wavefront aberrations, although
diffraction effect appears to be less of an issue in the current analysis.
Overall, the binning impact seems more prominent in the coefficients with 
odd radial orders ($k$=$1,3$) than in the even radial order coefficients ($k$=$2$).
The estimated slope coefficients still approximate the true values 
with sub-pixel accuracy. 

\noindent The maximum difference between $\hat{a}_{i}$ and $a_{i}$ 
is 0.032 wave in $a_{9}$. The through-focus PSFs after applying 
$\hat{a}_{i}$ as corrections are shown in the bottom row of 
Figure~\ref{Figure2}.
\begin{table}
  \centering
  \footnotesize
  \caption{\footnotesize{Slope (in pixel) and wave (in $\lambda$) coefficients in Figure~\ref{Figure2}.}}
  \begin{tabular}{c|c|c|c|c|c|c} \hline\hline
    $i^k$ & $\hat{a}{'}_i^x$ &  ${a}{'}_i^x$ & $\hat{a}{'}_i^y$ & $a{'}_i^y$ 
    & $\hat{a}_i$ & $a_i$\\ \hline
    1$^0$ &   -11.924 &   -11.959 &   -1.080  &  -1.144   &  --    &  --    \\ \hline
    2$^1$ &   -27.851 &   -27.986 &    7.928  &   7.968   & -1.182 & -1.185\\
    3$^1$ &     7.928 &     7.968 &  -30.884  &  -31.015   &  1.607  &  1.636\\ \hline
    4$^2$ &   -12.468 &   -12.505 &  -13.006  &  -13.316  & -4.239  & -4.241\\
    5$^2$ &     1.761 &     1.766 &  -18.385  &  -18.601  &  1.618 &  1.626\\
    6$^2$ &     0.753 &     0.917 &   20.155  &   20.598  &  0.310  &  0.309\\ \hline
    7$^3$ &     -- &     -- &    --  &   --  & -1.328  & -1.359\\
    8$^3$ &     -- &     -- &    --  &   --   & -1.273 & -1.276\\
    9$^3$ &     -- &     -- &    --  &   --   &  1.582  &  1.614\\
    10$^3$ &     -- &     -- &    --  &   --   &  1.381 &  1.408\\
\hline\hline
  \end{tabular}
  \label{Table2}
\end{table}

\begin{figure}[htpb]
	\centering 
 	\includegraphics[width=8.5cm]{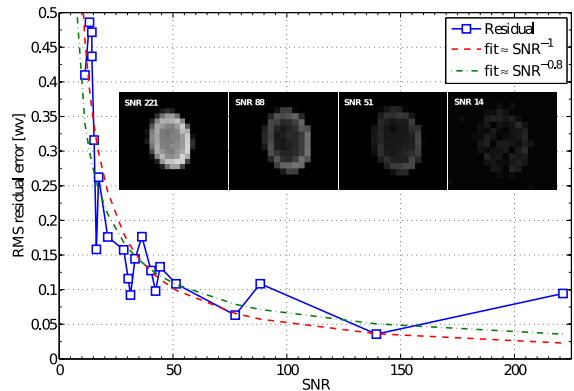}
	\caption{\footnotesize{Four object images at 1$\lambda$ FD (inset), 
	RMS residual error against SNR for 21 objects.}}
	\label{Figure3}
\end{figure}

Finally, random aberrations (from $a_2$ to $a_{15}$, i.e. M=15) were applied to 21 
point objects with different brightness. The aberration, without tip/tilt, amounts 
to 0.5$\lambda$ rms at 500nm. The images of four object are shown at the maximum FD 
of $1\lambda$ in the inset (Figure~\ref{Figure3}). Five FD frames were obtained, 
where the pixel size is 18$\mu m$ with scale of 0.15$''$ at 2.7m aperture and 
uniformly weighted 11 wavelengths between 500nm and 600nm were used with photon 
shot noise. The signal-to-noise ratio (SNR), the square root of total photons collected 
as computed at the zero FD, ranges from 11 
to 221. The rms residual error, i.e. the root sum square of $\hat{a}_i-a_i$, is 
plotted against SNR (blue) with two power-law fits.
While the green curve fits the blue data better, the red fit also asymptotes 
the data and follows the general behavior of the accuracy of other WFS techniques 
($\sim$SNR$^{-1}$)\cite{Hardy}. It appears that SNR higher 
than 50 per frame is needed to obtain accuracy better than 0.1$\lambda$. This 
translates to SNR$\sim$20 for a SHS with N=30 sub-apertures, based on the 
requirement of 2$M$$\leq$$N$\cite{DeSilva, Lofdahl}, leading to the theoretical SHS accuracy of 
0.05$\lambda$\cite{Hardy}. The analysis indicates the proposed technique 
to be suitable for high SNR cases, while spatial filtering, noise
reduction, or thresh-holding could help improve its robustness in low SNR cases.

The demonstrations suggest the feasibility of the proposed geometric concept in 
WFS with high SNR and strong aberrations in particular. The fact that the moment 
analysis is just an extension of the routine computation of image centroid or 
full-width-half-maximum can make all appropriate objects recorded in an
image frame be potential WFS targets. 
This can certainly permit a straightfoward and rapid WFS potentially 
over a wide field of view, which can be particularly attractive in 
active diagnosis of alignment-driven field aberrations of imaging systems\cite{Lee} 
or wide-field adaptive image compensation. The error propagation 
is to be separately discussed with other considerations including aberration 
aliasing, diffraction, and FD implementation.

\end{document}